\documentclass[a4paper,10pt]{article}
\pdfoutput=1
\usepackage[utf8x]{inputenc}
\usepackage[pdftex]{graphicx}
\usepackage{amssymb,amsfonts,amsmath}
\usepackage{setspace}

\usepackage{color,soul} 
\definecolor{violet}{rgb}{.93,.51,.93}
\sethlcolor{violet}

\newcommand{\qed}{\hfill $\Box$\\}
\newcommand{\R}{\mathcal{R}}

\newcommand{\Rz}{\mathrm{R}_0}

\title{Parameter Exploration in Simulation Experiments: A Bayesian Framework}
\author{{Jessica~W.~Leigh}\and{David~Bryant}}

\begin{document}

\maketitle

\begin{abstract}
\noindent Simulations often involve the use of model parameters which are unknown or uncertain. For this reason, simulation experiments are often repeated for multiple combinations of parameter values, often iterating through parameter values lying on a fixed grid. However, the use of a discrete grid places limits on the dimension of the parameter space and creates the potential to miss important parameter combinations which fall in the gaps between grid points. Here we draw parallels with strategies for numerical integration and describe a Markov chain Monte-Carlo strategy for exploring parameter values. We illustrate the approach using examples from phylogenetics, archaeology, and epidemiology.
\end{abstract}


%
%

\section{Introduction}

Simulation experiments are a fundamental tool of many scientific fields. They are used to make model comparisons, to predict future events, carry out sensitivity analysis and to promote and test hypotheses or methodologies. indeed simulation experiments are almost obligatory in papers introducing new methods or techniques. They can also be one of the most computationally intensive components of a scientific investigation, involving large scale calculations and a huge amount of replication. Simulation has also become an important tool for inferring parameters from data (e.g. \cite{Liu2008}), though this is not the application that we will be considering here.

In many fields, the models used in a simulation involve parameters which are either unknown or roughly estimated. One of the biggest practical challenges for those conducting simulations is deciding which values to use for these model parameters. The range of values needs to be broad enough that the experimental results have some level of generality, but still within the limits imposed by computational resources and time. A standard practice is to fix some values {\em a priori} and use a discrete grid to iterate over others. Multiple replicates are then carried out at each grid point (combination of parameter values).

Grid-based strategies for exploring a function or space are widely used in numerical integration. They typically face the curse of dimensionality: as the number of dimensions grows the number of grid points increases exponentially \cite{Liu2008}. 
This issue is, to an extent, avoided by Markov chain Monte-Carlo (MCMC) methods, where points are sampled randomly within the entire space according to an appropriately chosen Markov chain.

Despite the impact that MCMC methods have had on a wide range of scientific fields (e.g. \cite{Brooks2011,Diaconis2009,Liu2008}), they do not appear to have been applied to the implementation of simulation experiments themselves. 
A simulation strategy which samples parameter stochastically using a Markov chain, rather than iterating through a fixed grid, could be advantageous for several reasons: it bypasses, to an extent, the issues of dimensionality;  there is less scope for critical areas of the parameter space to fall `between the sample points'; and it may be possible to avoid some unnecessary simulations in `uninteresting' parts of parameter space by performing only one replicate per sample point. The MCMC algorithms samples more heavily from those area for which the outcome being studied has a higher probability, thereby mapping the surface more efficiently that, say, sampling from the prior alone.

There are several obstacles to overcome before MCMC methods can be used to conduct simulation experiments in this manner. First, MCMC is a technique for sampling from a distribution, and it is not immediately obvious what that distribution should be in this context. Second, MCMC and importance sampling algorithms typically require the value for a distribution to be known, at least up to a scalar constant. Given the model complexity inherent in many simulation experiments, exact or proportional values will not always be available.

We address the first obstacle by describing a Bayesian-style  framework for conducting simulation experiments in which the distribution of interest is the ``posterior'' distribution of the parameters conditional on a particular simulation outcome. We say ``Bayesian-style'' because we never work directly with actual data and the choice of prior is eventually factored out of the final output. We address the second obstacle using clever algorithms of \cite{Marjoram2003} and \cite{Didelot2011} for conducting MCMC and importance sampling in the absence of explicit likelihoods. Our methodology therefore overlaps with that used for {\em approximate Bayesian computing}, though we do not require the reduction of data to summary statistics. 
 
To illustrate our new framework, we compare two workhorses of phylogenetic inference, the UPGMA algorithm \cite{Sokal1963} and the Neighbor-Joining algorithm \cite{Saitou1987}. Both algorithms construct trees (or hierarchies) from distance data, the main difference between the two being that UPMGA assumes the mutations accumulate at a constant rate whereas NJ does not assume a constant rate, at the cost of a slightly higher sampling variance. We will use a simple simulation to identify situations where UPGMA performs at least as well as NJ, primarily to illustrate aspects of our MCMC framework.

We then describe two additional case studies. The first re-examines an important simulation experiment conducted by Allaby \cite{Allaby2008}. A key question of prehistoric agriculture is whether domesticated crops have single or multiple origins. Allaby et al \cite{Allaby2008} used simulations to demonstrate that standard phylogenetic based methods could infer a single origin even  given data from (simulated) crops with multiple origins. Aspects of Allaby et al.'s model have been contested \cite{Ross-Ibarra2008}. Here we consider the choice of parameters values, and ask whether the results of Allaby et al's study are robust to error in the model parameters. We conclude that they are robust, and use the example to illustrate how our MCMC approach can be used to explore multidimensional parameter spaces.

The final case study comes from epidemiology and uses a complex, and computationally intensive, simulator which models transmission of diseases through contact networks. In this study, we use population tract data for the city of Seattle. We consider two strategies for controlling the outbreak of a hypothetical disease: school closure and the provision of antiviral drugs. We examine how these strategies perform relative to each other, varying model parameters describing the rate of infection ($\Rz$) and the fraction of the population $f_v$ vaccinated prior to the start of the epidemic.

\section{Methods}

\subsection{Bayesian framework}

There are three principal ingredients of any simulation experiment: the simulator itself; the outcome under study; and the scheme used to select parameter values and replicate counts. The {\em simulator} takes a vector of parameter values $\theta$ and generates synthetic data $X$ according to a distribution $P\left(X|\theta\right)$ determined by the underlying model. Generally, the relationship between $X$ and $\theta$ is determined by an algorithm and actual values of $P(X|\theta)$ are unavailable.\\

\noindent {\bf Example}\\
{\em To compare NJ and UPGMA we simulate a ``true tree''  which is then used to evolve synthetic genetic data. The distribution of the tree is governed by two parameters: the ``scale'', $s$, which controls the height of the tree (and thereby the level of variability in the data) and the ``skew'', $\gamma$, which controls the variation in mutation rate in different parts of the tree. We used the sequence simulator implemented in the Python package P4 \cite{Foster2004} to evolve nucleotide sequences along this tree and then produced distance estimates from these. Details are in the appendix. }\qed

We assume that the goal of the simulation experiment is to study a particular {\em outcome}, which  corresponds to a particular event. For example, if we are using simulations to assess a given hypothesis test, the ``outcome'' of interest might be that the test gives a false positive and  the goal might be to identify parameter values for which the probability of this outcome, a false positive, is large. We let $\R$ denote the event that the outcome occurred, so $P(\R|\theta)$ is the probability of the outcome for a given set of parameter values $\theta$. \\

\noindent {\bf Example}\\
{\em For the NJ-UPGMA comparison the outcome $\R$  corresponds to the event that the tree produced by UPGMA is as close to the true tree as is the tree produced by NJ. We measure closeness using the Robinson-Foulds \cite{Robinson1981} distances between each of the UPGMA and NJ trees and the true tree. The outcome $\R$ can be any event, and could be quite complicated. Below we consider examples where $\R$ corresponds to the result of a hypothesis test, or where $\R$ corresponds to one strategy for mitigating an epidemic works better than another. }\qed

The probability $P(\R|\theta)$ for a particular value of $\theta$ is generally estimated by conducting multiple simulations with the same parameter values $\theta$ and recording the proportion of times $\R$ occurs. This is  computationally intensive, and limits the range of different $\theta$ values which can be considered. This is used in the standard `grid search' approaches for simulations:

\begin{enumerate}
\item[G1.] For all combinations of parameter values $\theta$
\item[G2.] \hspace{0.5cm} Carry out $r$ simulations using parameters $\theta$ (e.g. $r=100$)
\item[G3.]\hspace{0.5cm} Estimate $P(\R | \theta)$ by the proportion of  simulations for which $\R$ occurs.
\end{enumerate}

Our approach is to turn the conventional methodology on its head and follow a Bayesian strategy. We ask: ``if the simulation returns the specified outcome, what are the probable values of $\theta$?'', in other words, ``what is $P(\theta | \R)$?". For this distribution to make sense, we need to specify a prior distribution $P(\theta)$ on $\theta$. Bayes' rule then gives
\begin{equation}
P\left(\theta | \R\right) = P\left(\R|\theta\right) \frac{P\left(\theta\right)}{P\left(\R\right)}.\label{eq:Bayes}
\end{equation}
Note that the ``likelihood'' $P\left(\R|\theta\right)$ can (usually) not be directly evaluated. 

Samples from the posterior distribution $P(\theta | \R)$ indicate  parameter values for which the outcome is more likely. This can be useful in itself, but it doesn't indicate whether the outcome is particularly likely in the absolute sense.  For example, it may be that UPGMA tree is almost never as close to the true tree than the NJ tree and $P(\R | \theta)$ is small for all $\theta$.  However if the few times it is closer occur for the parameter values $\theta^*$ then $P(\theta^* | \R)$ will be large. 

In many contexts, it is more useful to infer the function $P(\R | \theta)$ as a function of $\theta$. This `likelihood' is also independent of the choice of prior. Below we show how to estimate this function using a combination of kernel density estimation and importance sampling.
%

\subsection{Sampling algorithm}

Our algorithm for sampling from $P(\theta | \R)$ is based on the ``MCMC without likelihoods'' algorithm of \cite{Marjoram2003}. Their algorithm takes data $\mathcal{D}$ and  generates samples from the posterior density $P(\theta|\mathcal{D})$. It does not use the likelihood function $P(\mathcal{D} | \theta)$ directly but instead simulates values  from $P(\cdot | \theta)$. Let $\pi$ denote the prior on $\theta$.
\begin{enumerate}
\item[F1.] If now at $\theta$ propose a move to $\theta^{\prime}$ according to a transition kernel $q(\theta \rightarrow \theta^{\prime})$. 
\item[F2.] Generate $\mathcal{D}^{\prime}$ using model $\mathcal{M}$ with parameters $\theta^{\prime}$.
\item[F3.] If $\mathcal{D} = \mathcal{D^{\prime}}$, go to F4, and otherwise stay at $\theta$ and return to F1.
\item[F4.] Calculate \[h = h(\theta,\theta^{\prime}) = \min \left(1,\frac{\pi(\theta^{\prime}) q(\theta^{\prime} \rightarrow \theta)}{\pi(\theta) q(\theta \rightarrow \theta^{\prime})} \right).\]
\item[F5.] Accept $\theta^{\prime}$ with probability $h$; otherwise stay at $\theta$; then return to F1.
\end{enumerate}
Marjoram and colleagues proved that the chain produced has the required stationary distribution, given an appropriate transition kernel $q$. 

To apply this algorithm within our context, we  replace the event $\mathcal{D} = \mathcal{D}^{\prime}$ with a generic event $\R$ (i.e, the result of a simulation experiment). We also switch the order of steps so that a simulation is carried out only if the move has not been rejected by the prior and transition kernel.

\begin{enumerate}
\item[S1.] If now at $\theta$ propose a move to $\theta^{\prime}$ according to a transition kernel $q(\theta \rightarrow \theta^{\prime})$. 
\item[S2.] Calculate \[h = h(\theta,\theta^{\prime}) = \min \left(1,\frac{P(\theta^{\prime}) q(\theta^{\prime} \rightarrow \theta)}{P(\theta) q(\theta \rightarrow \theta^{\prime})} \right).\]
\item[S3.] With probability $h$, go to S4; otherwise stay at $\theta$ and return to S1.
\item[S4.] Generate $X$ with distribution $P(X|\theta^{\prime})$ using the simulator and assess $\R$ using $X$.
\item[S5.] If $\R$ holds, accept $\theta^{\prime}$ and otherwise stay at $\theta$, then return to S1.
\end{enumerate}
Note that if the transition kernel is symmetric and the prior distribution is uniform then S2.~and S3.~can be deleted. 

With mild conditions on $q$, the chain produced by the algorithm converges to the required stationary distribution $P(\theta|\R)$, even though knowledge of the likelihood $P(\R | \theta)$ is never required. This property allows us to bypass the need for multiple replicates at each $\theta$ value. It also provides a way of sampling through the space of $\theta$ values without the need to resort to  grid sampling.\\

\begin{figure}[htbp]
\centerline{ \includegraphics[width=0.8\textwidth]{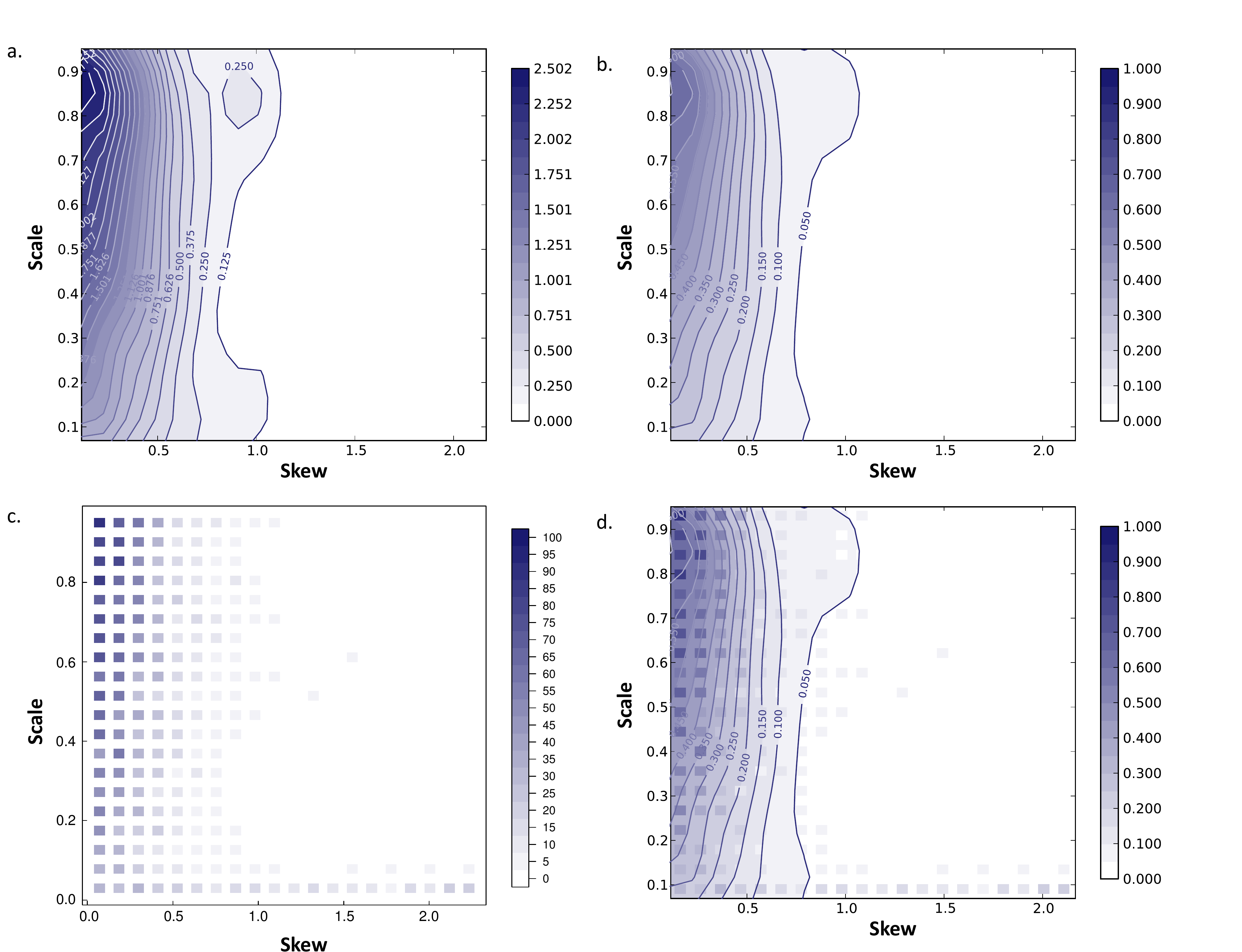}}
 \caption{\small Comparison of UPGMA and NJ. a: Probability density function for the distribution of skew and scale parameters, given that UPGMA performed at least as well as NJ; b: Likelihood function showing the probability that UPGMA performs at least as well as NJ; c: frequency with which UPGMA performed better than NJ over 100 simulations per parameter combination (i.e., a grid-search); d: plot c overlayed on plot b for ease of comparison. Darker colour indicates higher values.}
 \label{figs:upgmanj}
\end{figure}

\noindent {\bf Example}\\
{\em In figure~\ref{figs:upgmanj}(a) we give a density plot obtained from 100,000 iterations of the MCMC algorithm applied with a (bounded) uniform prior on $\theta_1$ and $\theta_2$. We see that if UPGMA performs at least as well as NJ, then it is likely that $\theta_1$ (the scale) is high and $\theta_2$ (skew) is small. This makes sense: UPGMA is known to perform better when the substitution rate does not vary across the tree (i.e., the tree is clocklike). It is also a lower variance method than NJ so will perform better as the data becomes noisier.} \qed

Samples from $P(\theta | \R)$ tell us which parameter values were likely, given that the simulation resulted in a specified outcome. For example, if a simulation was used to examine when a statistical test gave false positives, we could identify which parameter values were most likely to lead to a false positive result. However, the samples do {\em not} tell us how likely the false positives are for those parameter values. That is, samples from $P(\theta | \R )$ do not provide direct information about $P(\R | \theta)$. In the UPGMA-NJ simulation, we cannot tell from figure~\ref{figs:upgmanj}(a) whether UPGMA is a better method than NJ at any point.

From \eqref{eq:Bayes} we have that
\[P(\R | \theta) = \frac{P(\theta | \R) P(\R)}{P(\theta)}.\]
We assume that the prior $P(\theta)$ is known and estimate $P(\theta | \R)$ using a sample produced by the MCMC algorithm. The missing ingredient is $P(\R)$, the overall probability of the outcome when parameter values are sampled from the prior density. In general, the estimation of the normalising constant
\begin{equation}
P(\R) = \int_\theta P(\R | \theta) P(\theta) d \theta  \label{eq:Bayes2}
\end{equation}
is a challenging statistical and computational problem, even when the likelihood function $P(\R | \theta)$ is available **Citation: Gelman??**. Our approach, which follows \cite{Didelot2011} and uses importance sampling, is a computationally efficient approximation.

We assume that we have sampled a large number of approximately independent parameter vectors  $\theta_1,\theta_2,\ldots,\theta_n$  from $P(\theta | \R)$. When the dimension of the parameter space is not too large (say $\leq$ 10), a good approximation of the posterior distribution can be obtained using kernel density estimates (KDE), see \cite{Scott1992}. Let $K(\theta)$ denote the kernel density estimate. 

We sample $M$ new values $\theta^{(1)},\theta^{(2)},\ldots,\theta^{(M)}$ from $K$. For each we simulate a data set $X^{(i)}$ from $P(X | \theta^{(i)})$ and compute importance weights
\begin{equation}
 w_i = \left\{ \begin{array}{rl}
 \frac{P\left(\theta_i\right)}{K\left(\theta_i\right)} &\mbox{ if $\R^{(i)}$ holds} \\
  0 &\mbox{ otherwise}
       \end{array} \right.
 \label{equations:importance}
\end{equation}
The importance sampling estimate for $P(\R)$ is then 
\[P(\R) \approx \frac{1}{M} \sum_{i=1}^M w_i.\]
We note that repeating the entire estimation procedure for the {\em complement} of the outcome ${\R}^c$, that is, the outcome that $\R$ did {\em not} occur, we can estimate $P(\R^c)$. The difference between $P(\R) + P({\R}^c)$ and $1$ gives a test of the quality of estimation.

Given an estimate for $P(\R)$ we can substitute $K(\theta)$ for $P(\theta|\R)$ in \eqref{eq:Bayes2} and obtain an estimate for $P(\R | \theta)$.\\ 


\noindent {\bf Example}\\
{\em In the UPGMA-NJ example we calculated a $P(\R)$ value of $0.116$. We used this and the prior density to estimate the function $P(\R|\theta)$, 
Figure \ref{figs:upgmanj}(b), which is essentially a rescaling of the posterior density plot Figure \ref{figs:upgmanj}(a). As a comparison, Figure \ref{figs:upgmanj}(c) is an estimate of the same function carried out using a grid of $\theta$ values and performing 100 replicates of the simulation per grid point.} \qed


\subsection{Case study I: Assessing the robustness a simulation experiment}

The origin of agriculture was one of the defining moments of human history, and yet there is still debate over 
the nature and timing of the domestication process \cite{Allaby2008,Allaby2008a,Brown2009,Gross2010,Molina2011,Ross-Ibarra2008}. Whereas archaeobotanical evidence suggests a protracted and complex period of domestication, at least of Fertile Crescent cereals, phylogenetic evidence has suggested a single domestication origin. 
 Allaby and colleagues used a series of simulation experiments to demonstrate the phylogenetic techniques used could be misleading, and that, under a particular model of plant domestication, the admixture of two populations that emerged in independent domestication events can appear monophyletic \cite{Allaby2008,Allaby2008a} (see also \cite{Ross-Ibarra2008}).

 The model of domestication and admixture used in the simulations of \cite{Allaby2008} is governed by seven parameters controlling recombination, population sizes, and bottleneck durations. Four of the seven parameter values were fixed in advance using estimates from the literature, while the remaining three were set to a limited number of different values. Their experiment therefore demonstrates that the phylogenetic method is misleading for a {\em particular} choice of parameter values.  
We used our simulation framework to investigate how robust their conclusion was to variation in the simulation parameter values. 

The model used for these simulations is described in detail in \cite{Allaby2008a} (see also Figure~\ref{suppfigs:allabymodel}). Briefly, frequencies for twenty chromosomes were established for each of two ``wild-type'' populations of size $n_w$. In order to simulate a bottleneck, $n_b$ individuals  were then drawn (with replacement) from each of the wild-type populations, and this small population size was maintained for $t_b$ generations. In each new generation, when a new genome was created, each pair of chromosomes underwent a recombination event with probability $p_r$. Each population was then allowed to expand to $n_d$ individuals over the course of a single generation, and this new domesticated population size was maintained for $t_d$ generations, then the two domesticated populations were merged and the simulation continued for another $t_h$ generations with this admixed population (of size $n_d$). At the end of the simulation, a NJ tree was constructed from a Dice \cite{Dice1945} distance matrix 
estimated from two individuals sampled from each of the wild-type populations, the domesticated populations, and the admixed population (a total of ten individuals). If the inferred tree contained a split separating the individuals from the admixed population from the others, they were considered to appear (erroneously) monophyletic.

We performed simulations using the seven parameters described above $\theta = (n_w, n_b, t_b, n_d, t_d, t_h, p_r)$ and an ``outcome'' event $\R$ corresponding to the event that the phylogenetic method falsely inferred monophyly.  Prior densities for all parameters were uniform on intervals centred on the values used in \cite{Allaby2008}. We used the MCMC to sample parameter values $\theta$ from $P(\theta | \R)$, and then used the KDE and importance sampling techniques to estimate $P(\R | \theta)$ as a function of $\theta$. See the appendix for precise details.

See the appendix for precise details about the simulation experiment.

\subsection{Case study II: Comparing competing strategies for epidemics}

Our second case study compares two different strategies for mitigating epidemics. We use
the  FluTE software \cite{Chao2010}, which incorporates detailed demographic data to model contact networks and the spread of an epidemic through a population. In this instance, the population modelled is the city of Seattle. 

We considered two responses to the outbreak of an epidemic: closing schools and administering antiviral drugs. Using the simulator we can simulate the progression of the epidemic when the first strategy is implemented and then repeat the experiment with the same random seed but with the second strategy. 

At first, we let $\R$ be the outcome that the total number of symptomatic individuals is smaller when the schools were closed than when antiviral drugs were administered. We consider two model parameters, the 
basic reproduction number of the virus ($\Rz $) and the fraction of the population that was vaccinated prior to the start of the epidemic ($f_v$). The goal was to see which strategy worked best over different values for these parameters. 

We also consider another measure of success, the {\em peak} (rather than total) number of symptomatic individuals. To this end we repeated the analysis, this time with $\R$ denoting the event that the peak number of symptomatic individuals was smaller when schools were closed than when antiviral drugs were administered.

See the appendix for precise details about the simulation experiment. 

\section{Results}

\subsection{Case study I: Assessing the sensitivity of a simulation experiment}

\begin{figure}[htbp]
 \centerline{\includegraphics[width=\textwidth]{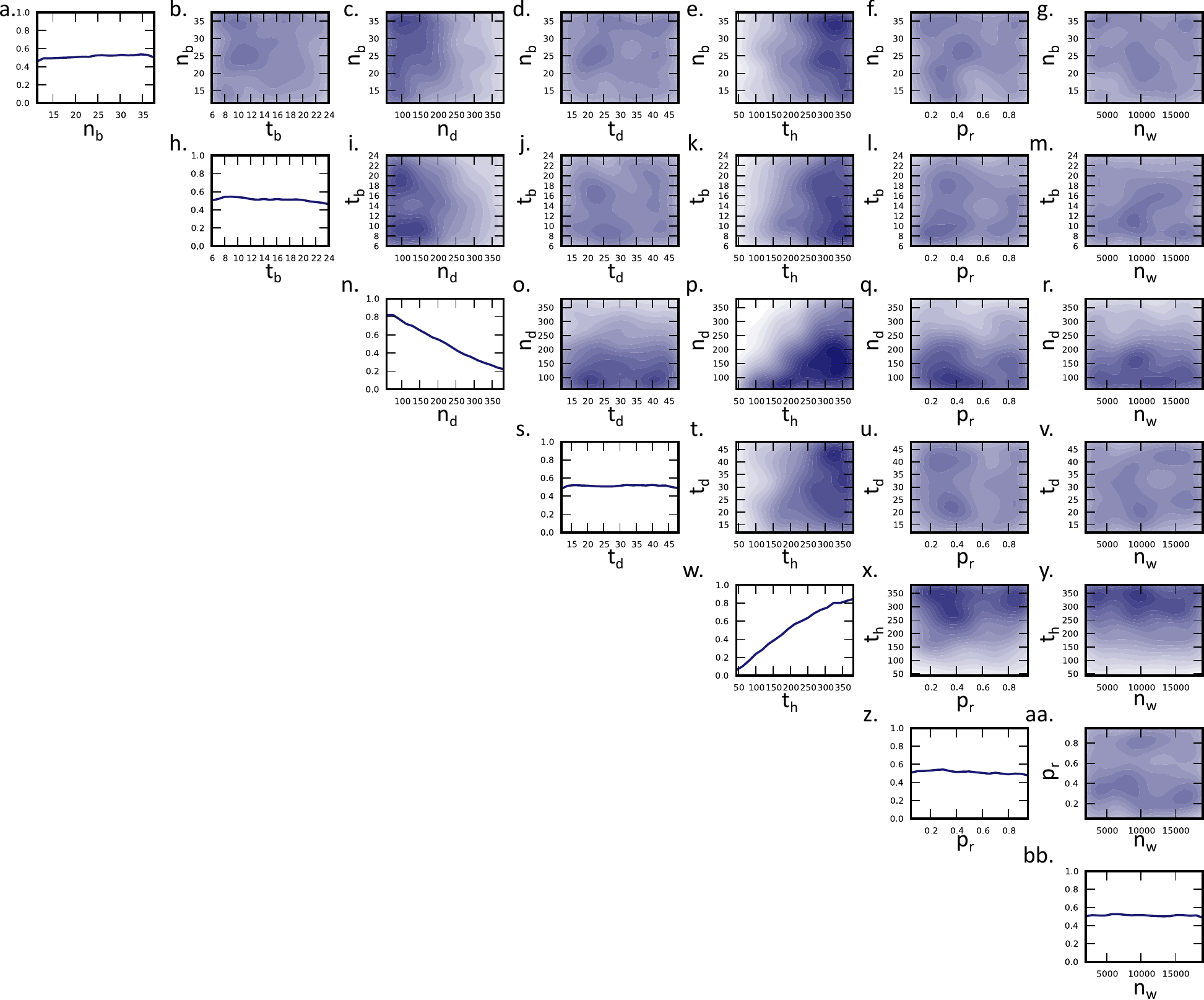}}
 \caption{\small Likelihood that a hybrid population erroneously appears to have a monophyletic origin. Likelihood functions showing the probability of inferring a monophyletic origin over single and pairs of parameters. $n_b$: size of the bottleneck populations; $t_b$: duration of the bottleneck (in generations); $n_d$: size of the domesticated populations (following post-bottleneck expansion as well as following hybridization); $t_d$: duration of domestication (prior to hybridization, in generations); $t_h$: time after hybridization (in generations); $p_r$: probability of a recombination event; $n_w$: wild-type population size. See text for a description of how these parameters were used.}
 \label{figs:domestication}
\end{figure}

Using the MCMC algorithm, kernel density estimation and the importance sampling method we obtained an estimate of the likelihood function $P(\R|\theta)$, which gives the probability of obtaining a (false) monophyly result for every choice of the six parameters. Figure~\ref{figs:domestication} gives the average likelihood values for every choice of single parameter and every pair of parameters. 

From the figure we see that variation in only two out of the six variables have a significant impact on the probability of monophyly: $n_d$, the  size of the domesticated population and $t_h$, the amount of time elapsing following admixture of the two domesticate populations. The probability of the test returning a false positive increases as the population size decreases or when the number of generations increases. Both trends are consistent with theory, as both decreasing the population size and increasing the number of generations  increases the chance that lineages from one of the domesticates have become fixed in the admixed population. A plot of probability of monophyly versus genetic diversity (Supplementary Figure~\ref{suppfigs:diversity}) suggests that, for these simulations, the diversity is a good predictor of a false positive test.

\begin{figure}[htbp]
 \centerline{\includegraphics[width=0.5\textwidth]{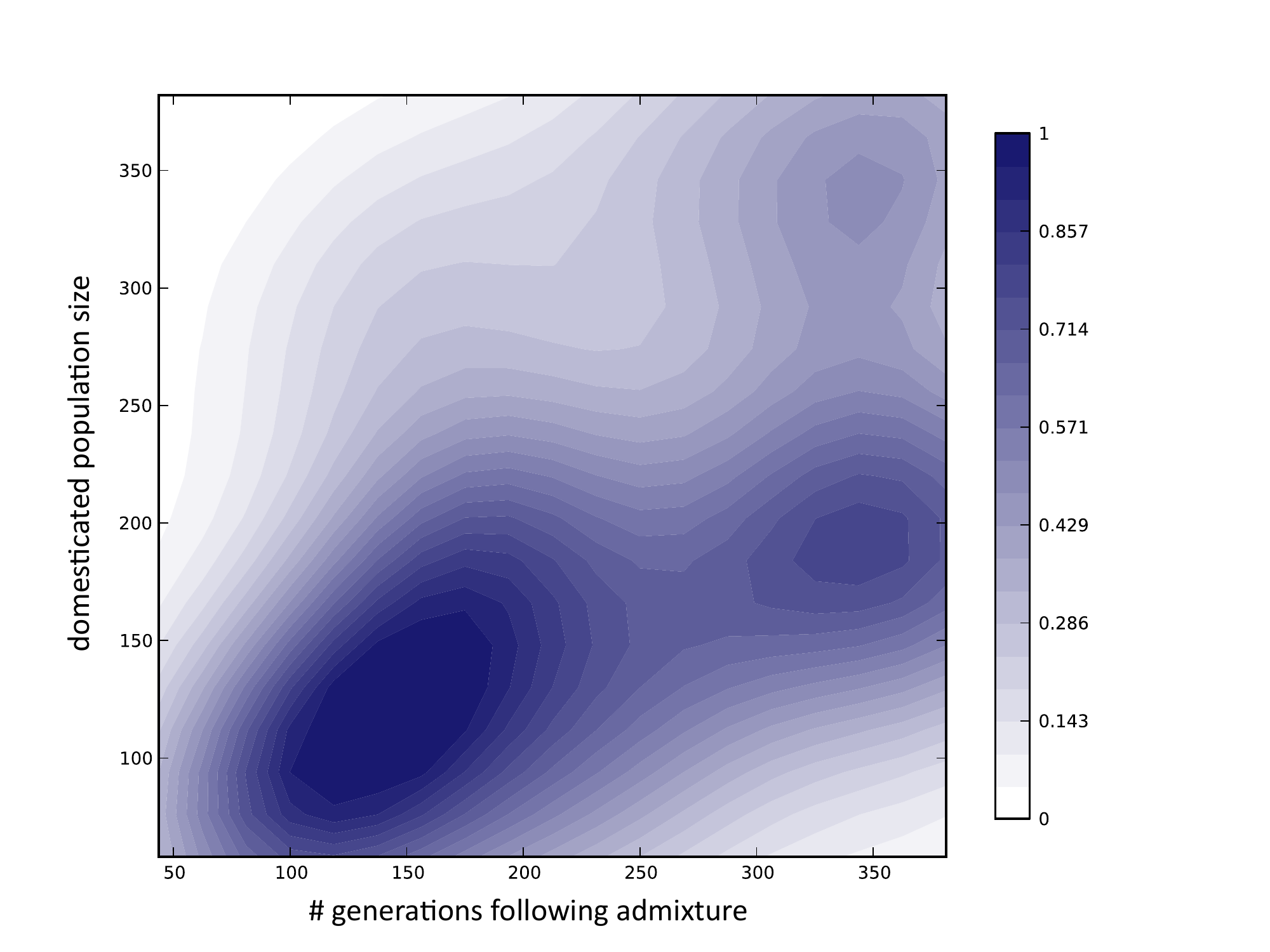}}
 \caption{\small Likelihood that a hybrid population appears monophyletic: Domesticated population size and post-admixture time. Likelihood function showing the probability of inferring a monophyletic origin estimated from the seven-parameter KDE by setting the remaining five parameters to the values used by Allaby and colleagues \cite{Allaby2008}, as described in the text.}
 \label{figs:ndtb}
\end{figure}

Figure~\ref{figs:ndtb} gives a more detailed representation of the probability of monophyly as a function of domesticated population size and bottleneck duration. Here we have fixed the other parameters at the same values used in the simulation of \cite{Allaby2008}, and  marked the various parameter values they used for the domesticated population size and the number of generations since admixture ($n_b = 20, t_b = 10, t_d = 20, n_w = 10000$). We fixed $p_r$, the recombination probability, at $0.1$, one of four values for this parameter used by Allaby and colleagues.

We see immediately that over the range of population size values considered, the probability of a false positive is generally high (\textgreater 0.5). Note, however, the strong dependency on the size of the domesticated population size. Ross-Ibarra and colleagues suggest that the small effective population sizes used in \cite{Allaby2008} are partially responsible for the results described \cite{Ross-Ibarra2008}. In response, Allaby and colleagues stated that the bottleneck parameters used in their model reflect established dimensions from biological populations \cite{Allaby2008a}.

\subsection{Case study 2: Comparing competing strategies for epidemics}

We applied our simulation framework to influenza epidemic simulation in order to assess the relative effectiveness of a non-pharmaceutical (school closure) and pharmaceutical (antiviral drugs) mitigation strategy. Simulations were set up using two different criteria to assess the success of a mitigation strategy over another: in one case, a strategy was considered successful if the {\em peak number} of symptomatic individuals was at most equal to the number when the other strategy was used; in the other, a strategy was successful if the {\em total number} of symptomatic individuals was at most equal to the number when the other strategy was used.

\begin{figure}[htbp]

\centerline{\includegraphics[width=0.5\textwidth]{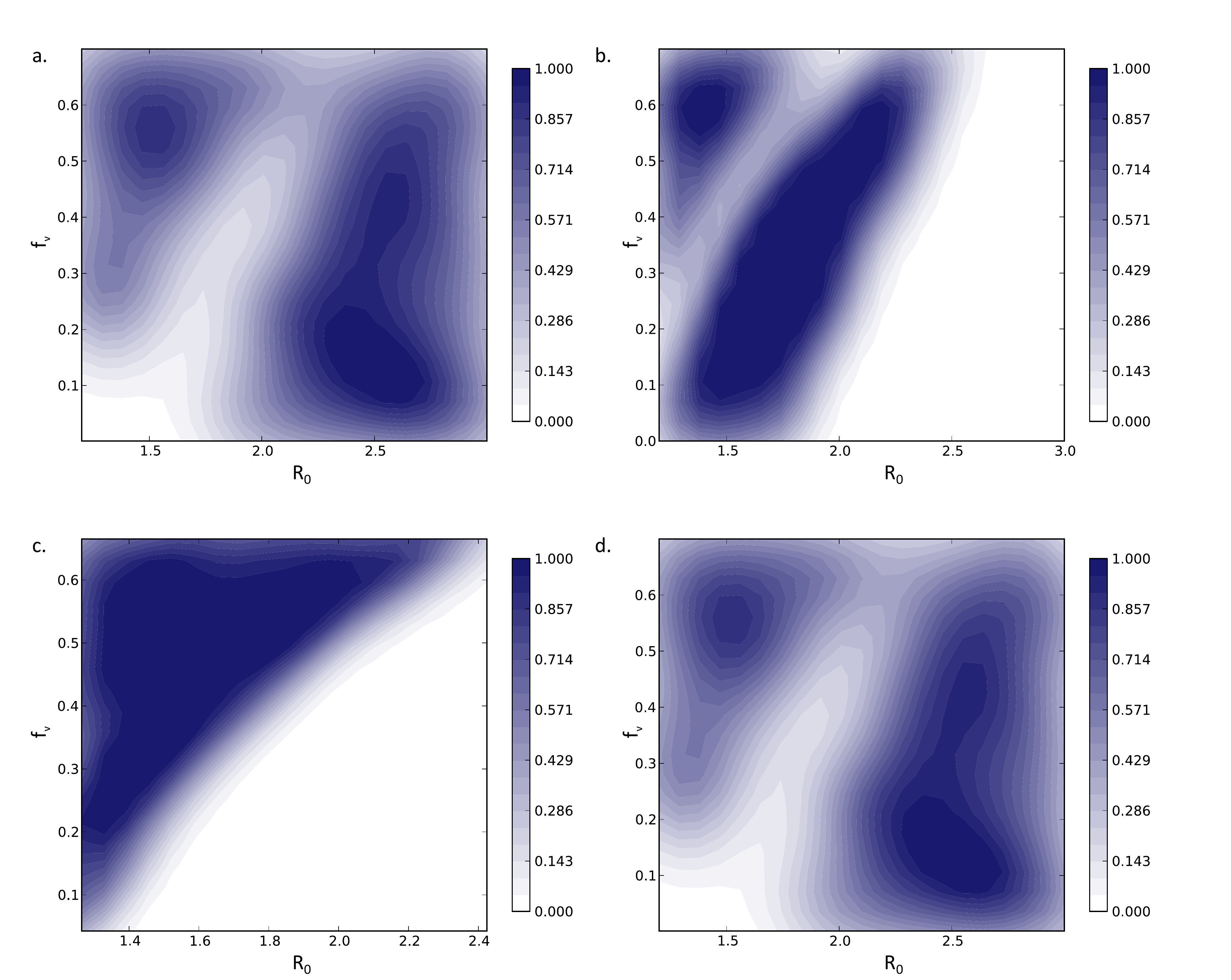}}

 \caption{Comparison of influenza epidemic response measures. Likelihood functions for simulated epidemics with two parameters, the basic reproductive number of the virus ($\Rz $), and the vaccinated fraction of the population ($f_v$). a: Probability that school closure reduces the peak number of cases to a greater extent than antiviral medication; b: Probability that antiviral medication reduces the peak number of cases to a greater extent than school closure; c: Probability that school closure reduces the cummulative number of cases to a greater extent than antiviral medication; d: Probability that antiviral medication reduces the cumulative number of cases to a greater extent than school closure.}
 \label{figs:flute}
\end{figure}

Figure~\ref{figs:flute}a and b show the distribution of the $\Rz $ and vaccinated fraction parameters sampled when success was based on the peak number of symptomatic individuals, when closure of schools or administration of antiviral drugs performed best, respectively. School closure (Figure~\ref{figs:flute}a) tended to reduce the peak number of cases relative to antivirals when either much of the population was vaccinated before the onset of the epidemic and the virus' $\Rz $ was low, or when little of the population was vaccinated and the $\Rz $ was high. The region corresponding to low to moderate $\Rz $ that was rarely sampled when school closure was the more effective strategy.  Antiviral drugs (Figure~\ref{figs:flute}b) also reduced the peak number of symptomatic individuals relative to closure of schools when the population was highly vaccinated and the $\Rz $ was low, but also when values for both parameters were small, moderate, or large. This 
corresponds almost exactly to the region that was rarely sampled when school closure was more effective than administration of antiviral medication.

Results when the acceptance criterion was a relative reduction in the total number of symptomatic individuals (Figure~\ref{figs:flute}c and d) differed greatly. In this case, school closure (Figure ~\ref{figs:flute}c) tended to be more effective than antiviral drugs across a broad range of $\Rz $ only when the population was highly vaccinated, but only for low $\Rz $ when the population was less highly vaccinated. In contrast, antiviral drugs (Figure ~\ref{figs:flute}c) were more effective than closure of schools across a broad range of $\Rz $ when the population was largely unvaccinated, and only for higher $\Rz $ values when the population was highly vaccinated.

Our FluTE simulations that used antiviral drugs as a response to the ascertainment of an influenza infection assumed that drugs were dispensed to both symptomatic individuals as well as their family members. FluTE models the effect of antiviral medications on an individual by decreasing three probabilities: the probability that a symptomatic individual will transmit the infection; the probability that an infected individual will become syptomatic in the first place; and the probability that an uninfected individual will become infected. Therefore, simulations using the antiviral strategy should show a reduction in the number of infections both within families of an infected individual and between members of an ``infected family'' and others.  When the population is considered as a contact network, with individuals as vertices and probability of infection as edge weights, it is clear that antiviral drugs decrease the weights of edges incident to infected individuals and their family members (who would have an 
increased risk of infection, due to increased contact probability with the infected individual).

On the other hand, school closure temporarily eliminates from the contact network those edges corresponding to within-school contacts, while at the same time increasing the weight of edges corresponding to daytime neighborhood contacts \cite{Chao2010}. Vertices corresponding to infected students will therefore generally have fewer incident edges, and if these edges largely correspond to vaccinated individuals, or if the probability of infection is generally low (i.e., if $\Rz $ is small), then many of these students may not infect others, and the total number of infections may be reduced. However, if it is very likely that an infected student will nonetheless infect at least one other person even when schools are closed (e.g., a neighbour or family member who has workplace contacts), and the infection can subsequently be transmitted to non-school environments with the same probability whether or not schools have been closed. The removal of within-school edges would therefore be expected to reduce 
the rate at which the epidemic will spread (and therefore the peak number), but not the total number of infected individuals.  These results 
are consistent with other studies, in which school closure has been shown to reduce the peak number of influenza infections to a greater extent than the total \cite{Chao2010, Ferguson2006, Kelso2009}. They are also consistent with the observation that school closure is less effective with higher $\Rz $ epidemics \cite{Halder2010}.

\section{Discussion}

Markov chain Monte Carlo algorithms have made a huge impact on numerical integration and statistical inference \cite{Liu2008}. Here we have presented a framework for conducting simulation experiments which uses MCMC to consider different combinations of parameter values. Our framework provides an alternative to a conventional `grid-based' strategy, where the simulation experiment iterates through all combinations of a fixed set of parameter values. 

The advantages (and disadvantages) of using a sampling based strategy to select parameter values are similar to the advantages (and disadvantages) of using Monte Carlo methods for numerical integration. Our approach is less vulnerable to the `curse' of dimensionality and, importantly, does not rely on {\em a priori} selection of fixed grid values, values which could completely miss regions of interest.

There are two main stages in the approach we describe. First, we use an MCMC algorithm to sample from a `posterior' distribution of the parameters conditional on a certain simulation outcome. For example, when comparing different strategies for dealing with epidemics we conditioned on the closing schools working better than administering antivirals, and then sampled from the distribution of two parameters: the reproduction number and the proportion of the population vaccinated. The resulting density indicates for which areas of the parameter space the school closure strategy is most likely to work better than administering antivirals.

The second stage takes the output of the MCMC and uses importance sampling to estimate the `likelihood' surface giving the probability of different outcomes for each parameter value. For example, in the epidemiological example the likelihood surface describes the probability that one strategy works better than another, whereas the output of the MCMC only indicates {\em where} one strategy was likely to work best. 

While we have demonstrated that the  framework we introduce is both feasible and practical, it is clear that considerable advances can be made improving the computational and statistical efficiency. The MCMC without likelihoods algorithm which we use is the basis of approximate Bayesian computation (ABC), and there has been a great deal of work improving the efficiency of ABC sampling algorithms \cite{Beaumont2009,Blum2010,Del-Moral2011}, most of which could be applied here. Our application of density estimation could likewise be improved and refined. Indeed, it is conceivable that there may be some way to use the simulation results from the MCMC step to sidestep the need for importance sampling, though we could not see how to do this without introducing a substantial bias.

We present the results from three case studies: a comparison of phylogenetic methods; a test of robustness for a simulation experiment in genetic archaeology; and an exploration of the effectiveness of two mitigation strategies for epidemics. In all three cases, our approach explored the parameter space without having to specify a fixed set of values. We obtained detailed estimates of the effect different parameter combinations had on the probability of the outcomes, estimates which could be used for further studies or to test hypotheses. 

Finally, we stress that the MCMC framework which we have introduced is extremely straightforward to implement. The basic sampler requires no more code than a grid based strategy, and the estimation of the likelihood function uses standard tools\footnote{Our implementation is available from {\tt http://leigh.net.nz/software.shtml}}. The framework is completely general and, while the applications explored here are all biological in nature, there is nothing to prevent the framework being applied in any context that simulation experiments are carried out.

\section{Acknowledgements}
JWL was supported by a postdoctoral fellowship from the National Science and Engineering Research Council (Canada). DB and JWL acknowledge support from the Allan Wilson Centre for Molecular Ecology and Evolution and the Department of Mathematics and Statistics, University of Otago.

\bibliographystyle{plain}
\bibliography{LeighBryant2012}

\pagebreak



\appendix

\section{Simulation details}

\subsection{Comparison of distance-based phylogenetic inference methods}

We compared two distance-based phylogenetic inference methods, UPGMA \cite{Sokal1963} and NJ \cite{Saitou1987}. Trees on thirty taxa were simulated under a Yule  model using the Python package DendroPy version 3.9 \cite{Sukuman2010}. A pair of parameters, $\theta$, describing the tree's non-clocklikeness (``skew''; $\gamma$) and expected root-to-tip distance (``scale''; $s$) were then used to adjust the edge lengths, as follows. For each edge in the tree, a uniform random number $u$ between $-\gamma$ and $\gamma$ was chosen, and the edge's length was multiplied by $\exp u$. The edges of the tree were then scaled so that the height of the tree was equal to $s$.

Nucleotide sequences of length 1000 were then simulated along this tree using the Python package P4 \cite{Foster2004} under the Jukes-Cantor model \cite{Jukes1969}. Pairwise distances between sequences were calculated using the Jukes-Cantor model, and trees were inferred using the NJ and UPGMA methods. The Robinson-Foulds \cite{Robinson1981} distance between each tree and the true tree (i.e., the tree along which sequences were simulated) was calculated.

We used this simulation strategy in the MCMC framework described above for 100,000 steps, saving every hundredth sample. Both parameters were drawn from uniform proposal distributions; $\gamma$ was uniform on $\left(0, \log 10\right)$, $s$ was uniformly distributed on $\left(0.02, 1\right)$. We then fitted a KDE to these 1,000 samples and used the importance sampling method described above to estimate the probability $P\left(\R\right)$ that UPGMA performs at least as well as NJ. One thousand additional simulations were performed in the application of the importance sampling method. 

We also used this simulation strategy in a grid-search approach by dividing each parameter in 20 intervals and performing simulations with the midpoints for all pairs of intervals (i.e., for all pairs of $\left(\gamma,s\right)$, where $$\gamma \in \left\{\frac{1}{40}\log 10, \frac{3}{40}\log 10, \dots, \frac{39}{40}\log 10\right\}$$ and $s \in \left\{0.0245, 0.0735, \dots, 0.9555\right\}$). For each pair $\left(\gamma,s\right)$, 100 simulations were performed, and the number of times UPGMA performed at least as well as NJ was counted.

\subsection{Investigation of erroneously inferred monophyly in domesticated populations}

Populations were simulated as described in \cite{Allaby2008}, implemented in custom software available from \hl{http://leigh.net.nz/software.shtml DO THIS}. A schematic of the model used in these simulations, described by seven parameters, is shown in Supplementary Figure~\ref{suppfigs:allabymodel}. Two sequences were sampled from each of the wildtype populations, from the two domesticated populations, and from the hybrid population. Dice distances \cite{Dice1945} were calculated between all pairs of sequences, and a tree was inferred using NJ \cite{Saitou1987}. If a bipartition was found in the inferred tree in which the two sequences from the hybrid population appeared on one side of the split to the exclusion of all other sequences, the population was considered monophyletic.

The probability distribution of the seven parameters, given that the hybrid population appeared monophyletic, was sampled using the MCMC method described here for 1,000,000 steps, sampling every 200 steps. All parameters were drawn from uniform proposal distributions with ranges given in Supplementary Table~\ref{supptabs:domest}. We then fitted a KDE to these 5,000 samples in order to apply the importance sampling method described here to estimate the probability $P\left(\R\right)$ of erroneously inferring a monophyletic origin for the hybrid population. For importance sampling, an additional 10,000 simulations were performed.

\subsection{Simulation of responses to an influenza epidemic}

Mitigation strategies for influenza epidemics were explored using the FluTE epidemic simulator version 1.13. \cite{Chao2010}. We used the population tract data for the city of Seattle, which is distributed with FluTE, for our simulations. Two variable parameters were explored in these simulations: the basic reproduction number of the virus ($\Rz $) and the fraction of the population that was vaccinated prior to the start of the epidemic ($f_v$). Other parameters used for simulations are summarized in Supplementary Table~\ref{supptabs:fluteparams}. Vaccinated individuals were distributed uniformly among age groups. For each combination of parameters, two simulations were run with the same seed (for pseudo-random number generation). When the number of symptomatic individuals reached the ascertainment threshold (0.8\%, see Supplementary Table~\ref{supptabs:fluteparams}), one of two possible mitigation strategies was initiated: either all schools were closed for a period of 14 days, or infected 
individuals and their household members were given antiviral drugs. The performance of the mitigation strategies was assessed based on either the peak number of symptomatic individuals, or the total number of symptomatic individuals. 

We used the MCMC framework described to explore the performance of the two mitigation strategies over the range of the two parameters ($\Rz $ and $f_v$) by sampling from the parameter space when school closure performed better than antiviral drugs, using each of the performance criteria (i.e., when either the peak or total number of symptomatic individuals was reduced relative to simulations in which antiviral drugs were administered). For comparison, we also sampled from the parameter space where antiviral medication was a more effective mitigation strategy using the relative reduction in both peak and total number of symptomatic individuals as a performance criterion. In each case, chains were run for 100,000 steps, sampling every 20 steps.  Proposal distributions for both parameters were uniform. $\Rz $ was bounded by $\left[1.2, 3\right]$ and $f_v$ was bounded by $\left[0, 0.7\right]$. We then used the importance sampling method described here to estimate the 
probability that each mitigation strategy outperforms the other under each of the two performance criteria, using 1,000 simulations in each case.

\section{Figure Legends}
\subsection*{Figure~\ref{figs:upgmanj}: Comparison of UPGMA and NJ}
a: Probability density function for the distribution of skew and scale parameters, given that UPGMA performed at least as well as NJ; b: Likelihood function showing the probability that UPGMA performs at least as well as NJ; c: frequency with which UPGMA performed better than NJ over 100 simulations per parameter combination (i.e., a grid-search); d: plot c overlayed on plot b for ease of comparison. Darker colour indicates higher values.

\subsection*{Figure~\ref{figs:domestication}: Probability of erroneously inferring a monophyletic origin for a hybrid population}
Likelihood functions showing the probability of inferring a monophyletic origin over single and pairs of parameters. $n_b$: size of the bottleneck populations; $t_b$: duration of the bottleneck (in generations); $n_d$: size of the domesticated populations (following post-bottleneck expansion as well as following hybridization); $t_d$: duration of domestication (prior to hybridization, in generations); $t_h$: time after hybridization (in generations); $p_r$: probability of a recombination event; $n_w$: wild-type population size. See text for a description of how these parameters were used.

\subsection*{Figure~\ref{figs:ndtb}: Probability of erroneously inferring monophyly: domesticated population size versus time following admixture}.
Likelihood function showing the probability of inferring a monophyletic origin estimated from the seven-parameter KDE by setting the remaining five parameters to the values used by Allaby and colleagues \cite{Allaby2008}, as described in the text.

\subsection*{Figure~\ref{figs:flute}: Comparison of influenza epidemic response measures}
Likelihood functions for simulated epidemics with two parameters, the basic reproductive number of the virus ($\Rz $), and the vaccinated fraction of the population ($f_v$). a: Probability that school closure reduces the peak number of cases to a greater extent than antiviral medication; b: Probability that antiviral medication reduces the peak number of cases to a greater extent than school closure; c: Probability that school closure reduces the cummulative number of cases to a greater extent than antiviral medication; d: Probability that antiviral medication reduces the cumulative number of cases to a greater extent than school closure.

\subsection*{Supplementary Figure~\ref{suppfigs:allabymodel}: Model from \cite{Allaby2008a} used for plant domestication simulation experiments}
Simulations were performed by sampling bottleneck populations from each of two separate wildtype populations. Following the bottleneck period, each bottleneck population expanded in size to form a domesticated population. Following the period of domestication these populations were then merged to form the hybrid population. The simulation continued for the duration of the hybridization period.

\subsection*{Supplementary Figure~\ref{suppfigs:diversity}: Relationship between erroneous inference of monophyly and genetic diversity}
The probability of inferring monophyly was plotted versus the genetic diversity of a simulated hybridized population for 1,000 sets of parameters randomly selected from within the range of the prior. Probabilities were estimated from the likelihood function shown in Figures~\ref{figs:domestication} and \ref{figs:ndtb}.




\clearpage

\section*{Supplementary Figures}
\setcounter{figure}{0}
\renewcommand{\thefigure}{S\arabic{figure}}

\begin{figure}[ht]
 \includegraphics[width=0.5\textwidth]{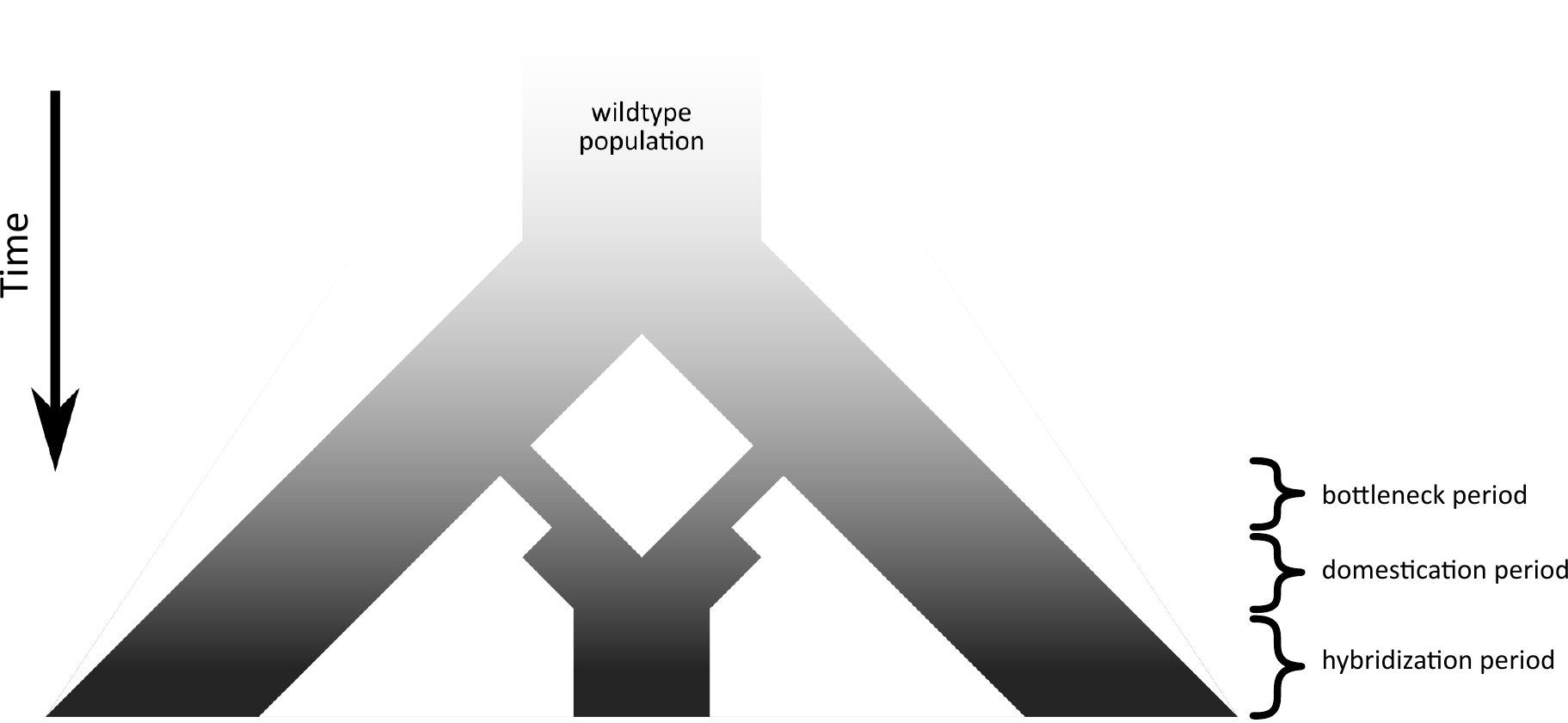}
 \caption{Model from \cite{Allaby2008a} used for plant domestication simulation experiments.}
 \label{suppfigs:allabymodel}
\end{figure}

\begin{figure}[ht]
 \includegraphics[width=0.5\textwidth]{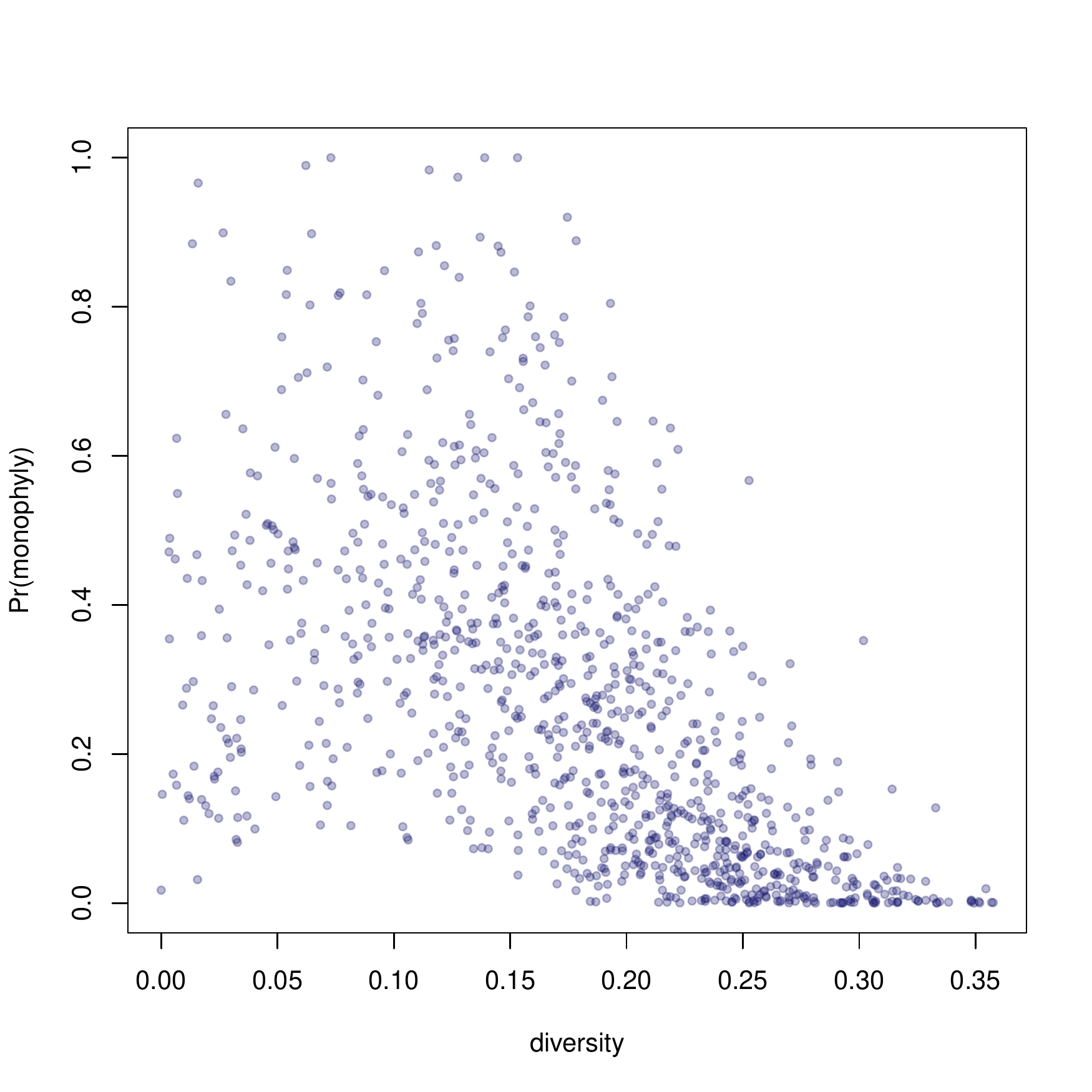}
 \caption{Probability that the hybrid population appears monophyletic as a function of genetic diversity}
 \label{suppfigs:diversity}
\end{figure}



\clearpage
\section*{Supplementary Tables}
\setcounter{table}{0}
\renewcommand{\thetable}{S\arabic{table}}

\begin{table}[ht]
 \caption{Parameters used in all FluTE simulations\textsuperscript{\textdagger}}
 \label{supptabs:fluteparams}
 \begin{tabular}{@{\vrule height 10.5pt depth4pt  width0pt}llll}
  \multicolumn{4}{l}{}\\
  \multicolumn{4}{l}{\textbf{a. Pharmaceutical efficacy parameters}}\\
  \textbf{Pharmaceutical} & \textbf{Susceptibility} & \textbf{Infectiousness} & \textbf{Illness, given infection} \\
  Vaccine & 0.4 & 0.4 & 0.67 \\
  Anti-viral & 0.3 & 0.62 & 0.6 \\
  \\
  \multicolumn{4}{l}{\textbf{b. Other parameters}}\\
  \textbf{Parameter} & \multicolumn{3}{l}{\textbf{Value}} \\
  Infected (initial) & \multicolumn{3}{l}{10} \\
  Infected (seeded daily) & \multicolumn{3}{l}{1} \\
  Simulation duration & \multicolumn{3}{l}{180 days} \\ 
  Ascertainment fraction & \multicolumn{3}{l}{0.8} \\
  Ascertainment delay & \multicolumn{3}{l}{1 day} \\
  Response threshold & \multicolumn{3}{l}{0.01}\\
  Response delay & \multicolumn{3}{l}{7 days} \\
  Pre-epidemic strategy & \multicolumn{3}{l}{vaccination (uniform)} \\

 \end{tabular}

  \textsuperscript{\textdagger} For parameters not listed, default values were used.\\
\end{table}

\begin{table}[ht]
 \caption{Bounds on parameters used in simulations based on the model of Allaby et al \cite{Allaby2008a}}
 \label{supptabs:domest}
 \begin{tabular}{@{\vrule height 10.5pt depth4pt  width0pt}lll}
  \textbf{Parameter} & \textbf{Minimum} & \textbf{Maximum}\\
  $n_w$ & $1000$ & $20000$\\
  $n_b$ & $10$ & $39$\\
  $t_b$ & $5$ & $25$\\
  $n_d$ & $40$ & $400$\\
  $t_d$ & $10$ & $50$\\
  $t_h$ & $25$ & $400$\\
  $p_r$ & $0$  & $1$\\

 \end{tabular}

\end{table}


\end{document}